\documentclass[english,notitlepage, twocolumn]{revtex4-1}
\usepackage[T1]{fontenc}
\usepackage[latin9]{inputenc}
\usepackage{color}
\usepackage{amsmath}
\usepackage{amssymb}
\usepackage{graphicx}

\usepackage{babel}
\begin{document}

\title{Negative refraction in time-varying, strongly-coupled plasmonic antenna-ENZ systems}

\author{V. Bruno$^{1,\dagger}$, C. DeVault$^{2,3,\dagger}$, S. Vezzoli$^{4,\dagger}$, Z. Kudyshev$^{2,3}$, T. Huq$^{4}$, S. Mignuzzi$^{4}$, A. Jacassi$^{4}$, S. Saha$^{2,3}$, Y.D. Shah$^{1}$, S.A. Maier$^{4,7}$, D.R.S. Cumming$^{6}$, A. Boltasseva$^{2,3}$, M. Ferrera$^{5}$, M. Clerici$^6$, D. Faccio$^{1,*}$, R. Sapienza$^{4,*}$, V.M. Shalaev$^{2,3,*}$}

\affiliation{$^1$ School of Physics and Astronomy, University of Glasgow,  G12 8QQ Glasgow, United Kingdom}
\affiliation{$^2$ Purdue Quantum Science and Engineering Institute, Purdue University 1205 West State Street, West Lafayette, Indiana 47907, USA}
\affiliation{$^3$ School of Electrical and Computer Engineering and Birck Nanotechnology Center, Purdue University, 1205 West State Street, West Lafayette, Indiana 47907, USA}
 \affiliation{$^4$ The Blackett Laboratory, Department of Physics, Imperial College London, London SW7 2BW, United Kingdom}
\affiliation{$^5$ Institute of Photonics and Quantum Sciences, Heriot-Watt University,  EH14 4AS Edinburgh, United Kingdom}
\affiliation{$^6$ School of Engineering, University of Glasgow, G12 8LT Glasgow, United Kingdom}
\affiliation{$^7$ Chair in Hybrid Nanosystems, Faculty of Physics, Ludwig-Maxilimians-Universitat Munchen, 80799 Munchen, Germany}

\email{ daniele.faccio@glasgow.ac.uk, r.sapienza@imperial.ac.uk, shalaev@purdue.edu: $^{\dagger}$These authors contributed equally.}

\begin{abstract}
Time-varying metasurfaces are emerging as a powerful instrument for the dynamical control of the electromagnetic properties of a propagating wave. Here we demonstrate an efficient time-varying metasurface based on plasmonic nano-antennas strongly coupled to an epsilon-near-zero (ENZ) deeply sub-wavelength film. The plasmonic resonance of the metal resonators strongly interacts with the optical ENZ modes, providing a Rabi level spitting of $\sim 30\%$.  Optical pumping at frequency $\omega$ induces a nonlinear polarisation oscillating at $2\omega$ responsible for an efficient generation of a phase conjugate and a negative refracted beam with a conversion efficiency that is more than four orders of magnitude greater compared to the bare ENZ film. The introduction of a strongly coupled plasmonic system therefore provides a simple and effective route towards the implementation of ENZ physics at the nanoscale.
\end{abstract}
\maketitle

{\bf{Introduction.}}
Time-varying systems and   metasurfaces are of interest in view of the fundamental physics questions that have arisen  \cite{mendoncca2000theory, nation2012colloquium,yablonovitch1989accelerating, westerberg2014experimental, PRLprain, bacot2016time, VezzoliTime}  
and also in view of the potential applications ranging from perfect lenses to spectral and temporal shaping of light fields \cite{Trety,pendry2008time, shaltout2015time, sounas2017non, shaltout2019spatiotemporal,sartorello2016ultrafast,nicholls2017ultrafast,rahmani2018nonlinear,nicholls2019designer,ciattoni2018phase}. Recent results have shown that thin films of  epsilon-near-zero (ENZ) materials with a dielectric permittivity close to zero \cite{liberal2017ENZreview,alu2007epsilon} at optical wavelengths in the visible or near-infrared spectral regions are promising candidates to achieve rapid (on the optical wave oscillation timescale) temporal changes of the optical properties \cite{VezzoliTime}. The very large order-of-unity refractive index changes that can be induced optically \cite{alam2016ENZnonlinear,PRLCaspani,clericiNC,carnemolla2018degenerate} makes it possible to achieve efficient temporal modulation uniformly across the medium \cite{shaltout2015time,marini2016self} even in deeply subwavelength thin films \cite{capretti2015TGHITOENZ,luk2015ENZmodes,capretti2015THGSicompENZ}, resulting in optically-induced negative refraction 
with unity efficiency \cite{VezzoliTime}. However, the results demonstrated so far rely on high-intensity optical pumping of the ENZ film in order to achieve such large changes in the refractive index. Recently, the combination of ENZ films with plasmonic structures has led to a significant reduction of the required optical powers for the Kerr nonlinear contribution to the refractive index \cite{Boyd2018large}.\\
Coupling between light and matter can be enhanced when two resonant systems with the same optical resonant frequency are brought into close contact \cite{torma2014strongReview}.  Strong coupling occurs when the strength of the coupling mechanism (measured by the splitting of the two resonant frequencies \cite{deliberato:timeModulatedVacuumRabifrequency}) dominates the intrinsic losses in the system thus resulting in a double peaked structure in the absorption spectrum or equivalently, in two well-separated polariton branches in the spectral domain. In the temporal domain, this will give rise to Rabi oscillations between the populations on these two branches and the combination of light-matter states where the matter component can contain a large fraction of the total energy. Strong coupling has been observed in a variety of systems  \cite{Brener2013strong, torma2014strongReview}, ranging from single atoms in cavity \cite{thompson1992observation}, quantum dots in photonic crystal \cite{yoshie2004vacuum} to Bose-Einstein condensates \cite{plumhof2014room} and superfluids \cite{amo2009superfluidityCarusotto}.\\ 
Strong coupling  at room temperature has also been reported between plasmonic resonators and deeply sub-wavelength ENZ films  \cite{jun2013ENZstrong,Boyd2016ENZpiuAnt,campione2016ENZstrong,gapPlasmonENZ2018coupling,passler2018strong}. 
 In this strongly coupled system, the fundamental plasmonic resonance of a metal antenna resonator is coupled to optical modes supported by the deeply sub-wavelength ENZ thin film at the frequency where the real part of the dielectric permittivity crosses zero, called ENZ modes. The ENZ modes can be seen as a long-range surface waves which arise from the interaction of two Surface Plasmon Polaritons (SPPs) at the two interfaces of the thin film~\cite{vassant2012ENZmodes,campione2015ENZmodes,campione2016ENZmodes2,runnerstrom2017ENZmode,luk2015ENZmodes,Vassant:12}. These ENZ modes exhibit a large density of states and can homogeneously confine the EM radiation within the ENZ \cite{campione2015ENZmodes,PhysRevA.87.053853}. Due to the impedance mismatch at the interface between air and the ENZ medium, excitation of the ENZ optical modes is inefficient, while adding strongly coupled antennas enables high electromagnetic  fields inside the ENZ film, resulting in enhanced nonlinear responses. \\ 
\begin{figure}[t!]
\centering
\includegraphics[width=8cm]{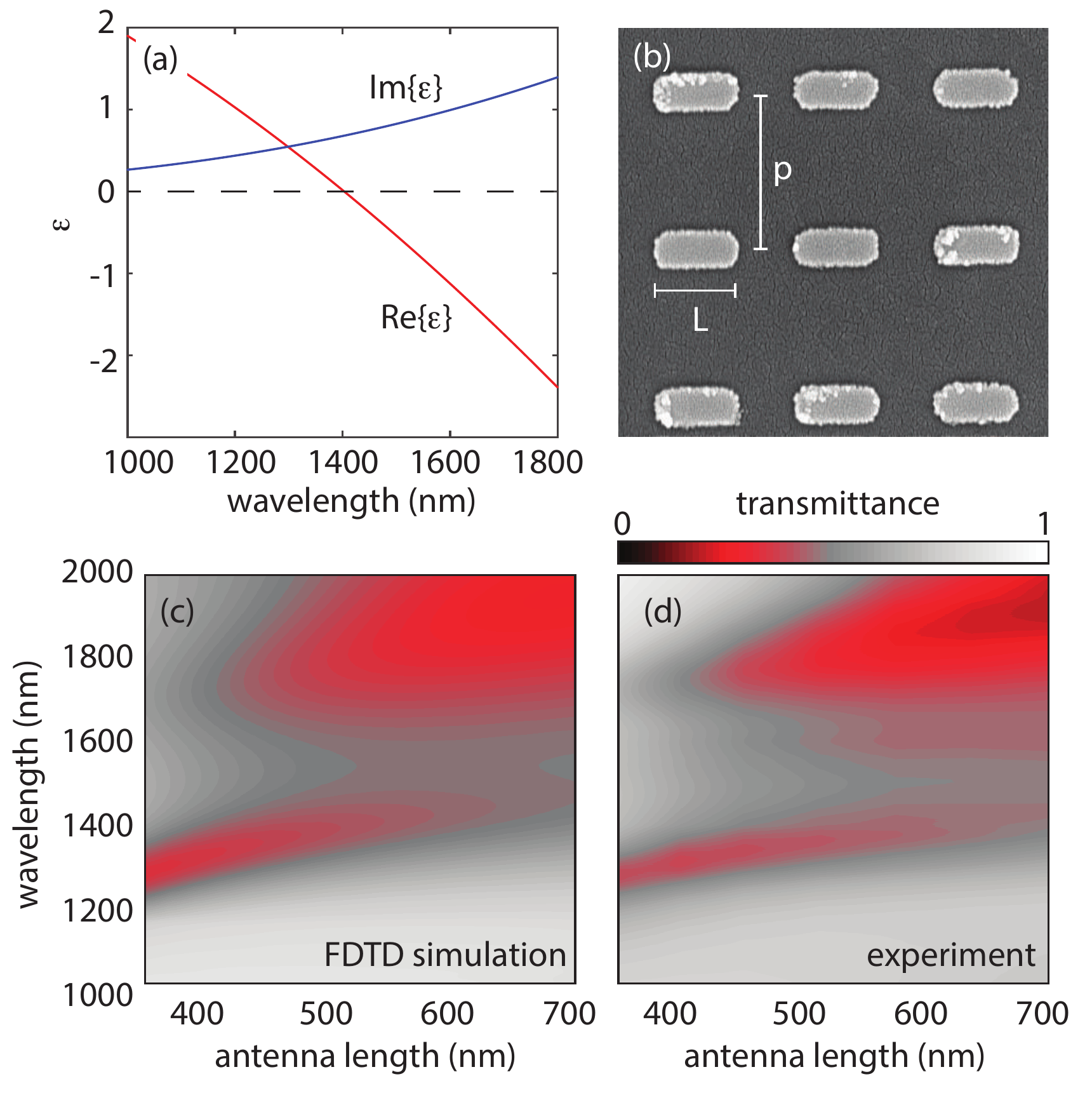}
\caption{ \textbf{Linear properties}. (a) Dielectric permittivity ($\varepsilon$) for the ITO layer based on a Drude model with real part Re$\{\varepsilon\}$ and imaginary part Im$\{\varepsilon\}$. (b) SEM image of the square lattice antenna pattern on ITO film (gold antenna length L = 460 nm, with period $p=800$ nm on a $d=40$ nm thick ITO film). (c) FDTD calculations of linear transmission through the metasurface using the ITO dielectric permittivity in (a). (d) Experimentally measured linear transmission of the metasurface (linear interpolation of measurements performed for antenna lengths of 400, 450, 500, 600, and 650 nm, $p=800$ nm). }
\label{fig:strongcoupling}
\end{figure}
Here, we study optically-induced negative refraction from a time-varying, strongly-coupled ENZ metasurface based on gold nano-antennas on top of a deeply sub-wavelength ENZ film. Experiments were performed by optically pumping at normal incidence a metasurface composed of rectangular metallic nano-antennas on a 40-nm-thick ENZ film  and probing the resulting temporal variation of the metasurface with a probe beam incident at a small angle. The generation efficiency of the negative refracted (NR)  and phase conjugated (PC) waves can provide a quantitative estimate of the strong coupling between the ENZ and the antenna modes, resulting in efficient optically-induced temporal variations of the material properties across a broad bandwidth (1200  - 1700 nm). The optically-induced temporal modulation \cite{Trety,pendry2008time} and resulting negative refraction and phase conjugation of the input probe beam generated in the strong-coupling regime are four orders of magnitude larger and cover a bandwidth that is three times broader in the ENZ wavelength region when compared to the bare ENZ film. \\
{\bf{Metasurface properties.}} Figure~\ref{fig:strongcoupling}(a) shows the real and imaginary parts of the dielectric permittivity via ellipsometry measurements of the 40 nm ITO film on a 1-mm-thick SiO$_2$ substrate. The ITO film exhibits a zero-crossing of the real part at 1400 nm (ENZ wavelength).  Figure~\ref{fig:strongcoupling}(b) is an example SEM image of the metasurface (top view) showing the geometry of the gold antennas deposited on the ITO surface with length L and periodicity p (see Supplementary information). 
The length of the antennas is such that their plasmonic resonance crosses the ENZ wavelength  of the ITO film and the periodicity p~$\sim600-800$ nm of the square lattice was chosen so as to maximise the density of the antennas whilst avoiding antenna-to-antenna coupling.\\
Figures~\ref{fig:strongcoupling}(c) and (d) show the numerically simulated (FDTD) and measured transmission spectra around the ENZ wavelength. For an incident optical beam normal to the metasurface and polarized along the long axis of the antenna, the spectra show two resonances corresponding to the two polariton branches of the strongly coupled metasurface.  The experimental spectral splitting of the two polariton branches is $\sim 420$ nm corresponding to a strong coupling efficiency of $\sim32$\% (see Supplementary information) \cite{campione2015ENZmodes}. \\
%
 \begin{figure}[t!]
\centering
\includegraphics[width=8.5cm]{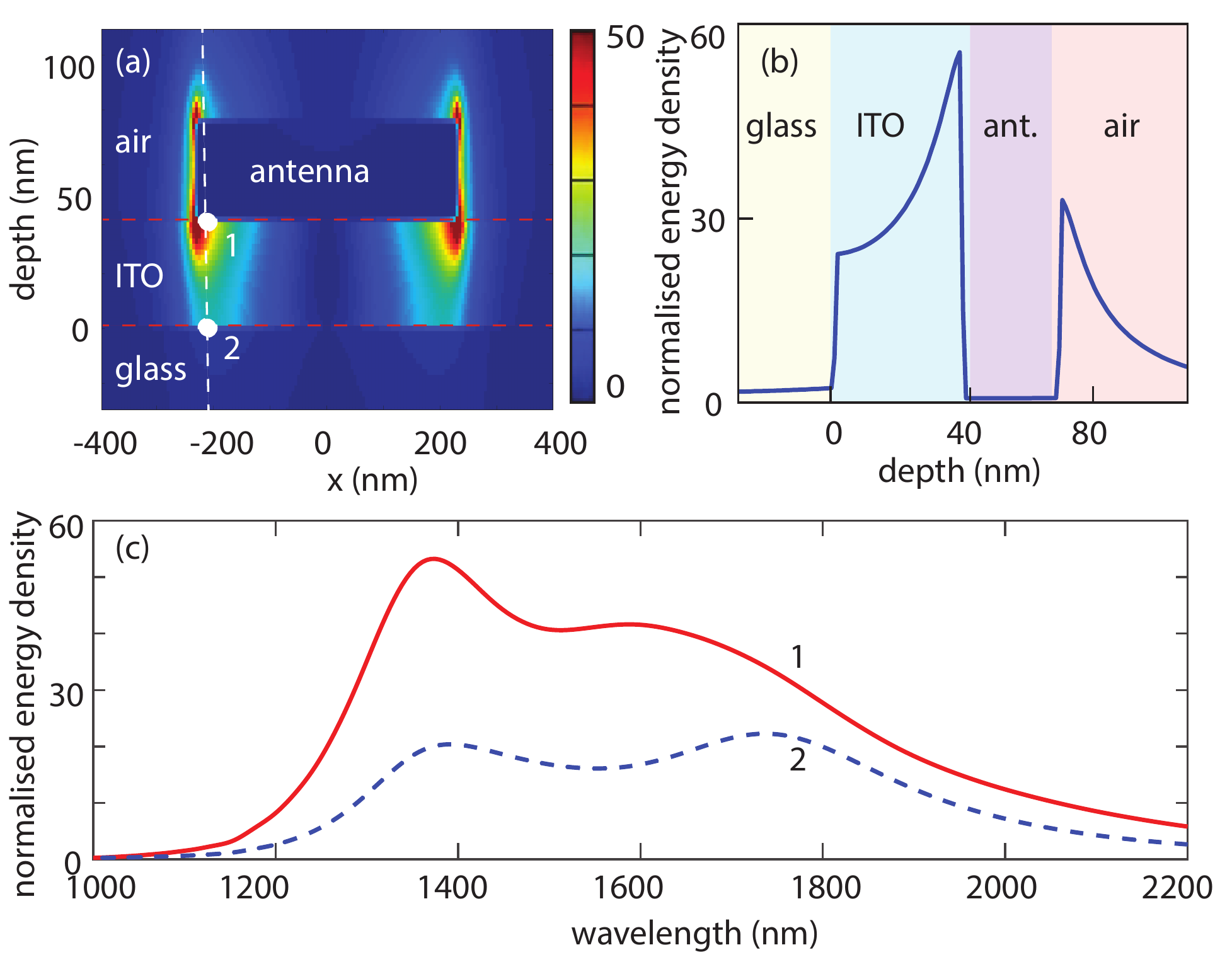}
\caption{ \textbf{FDTD simulation} of the normalised energy density (calculated as the $|E|^2$-field distribution normalised to the input $|E|^2$) in (a) for the 2D nano-antennas pattern on top of a 40 nm layer of ITO. (b) Depth profile of the normalised energy density across the metasurface [dashed white line in (a)]. (c) Shows the normalised energy density  versus wavelength calculated at the two points indicated with ``1'' and ``2'' in (a). }
\label{fig:field}
\end{figure}
In Fig.~\ref{fig:field}(a) we report the normalised energy density, calculated from finite difference time domain (FDTD) simulations as the $|E|^2$-field distribution normalised to the input $|E|^2$ distribution  at 1400 nm wavelength for the metasurface (L = 460 nm and p = 800 nm) upon normal incidence of the pump beam. Figure~\ref{fig:field}(b) shows the normalised energy density along the vertical line [white dashed line in (a)]. For a 1400 nm laser pulse polarized parallel to the long axis of the antenna and at normal incidence, the field intensity is enhanced by a factor greater than 50 with respect to the bare ITO.  
In Fig.~\ref{fig:field}(c) we plot the wavelength dependence of the energy density across the full bandwidth covering the two polariton branches calculated at two different points indicated as ``1'' and ``2'' in Fig.~\ref{fig:field}(a), showing an enhancement that is  40 times greater with respect to the bare ITO layer over a $\sim300$ nm bandwidth. \\
{\bf{Experiments.}}
\begin{figure}[t!]
\centering
\includegraphics[width=8cm]{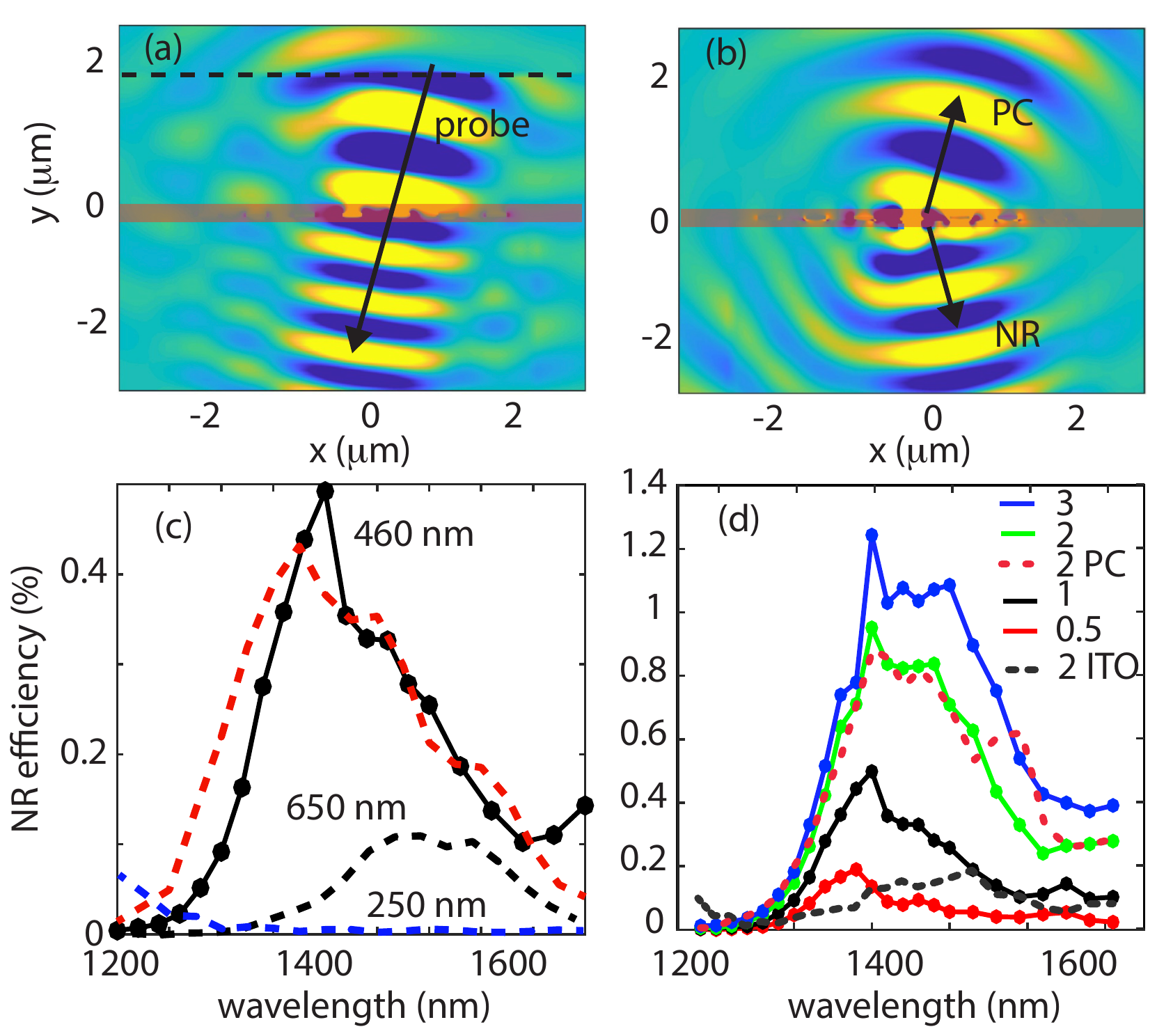}
\caption{ \textbf{Simulations and experiments}. (a) Nonlinear FDTD simulation of input probe beam. The arrow indicates probe direction (pump not shown), the red shaded area shows the metasurface area, and the dashed line indicates location of E-field source. (b) Phase conjugated (PC) and negative refracted (NR) beam are generated at the metasurface.  (c)  
	NR efficiency (NR energy normalised to input probe energy, expressed in $\%$) simulated with a nonlinear FDTD based on the same material and geometry as those used in the experiments (dashed red line) and antenna length $L=460$ nm (indicated in the graph). Also shown (solid-black curve) is the measured curve for 1 GW/cm$^2$. The blue and black dashed lines show the simulated NR efficiency with plasmonic antennas with longer or shorter lengths  (250 and 650 nm, indicated in the graph) so as to be detuned from the ENZ wavelength. The NR efficiency drops by nearly an order of magnitude thus highlighting the role played by strong coupling in the FWM enhancement.
	(d) The measured negative refraction (and an example of PC) signal efficiency for antennas on a film of 40 nm of ITO  (antenna length $L=460$ nm, period $p=800$ nm) for various pump powers indicated in the graph legend in GW/cm$^2$. The NR efficiency of the bare ITO film for a pump power of 2 GW/cm$^2$, multiplied by 1000 is shown for reference (dashed black curve). }
\label{fig:results}
\end{figure}
By using a pump and probe set-up, we perform a degenerate four wave mixing (FWM) experiment (i.e. a single pump beam and a single probe beam, both at the same wavelength) in the 1180 nm to 1710 nm spectral range. The optical pump beam has normal incidence on the sample, while the probe is incident at a small ($6^\circ$) angle. The two incident laser pulses are co-polarized (parallel to the long axis of the antenna) and have a temporal duration of 240 fs, the same central wavelength, and 100 kHz repetition rate. The generated NR and PC  are measured with a photodiode and compared to the transmission of the bare ITO in order to evaluate the efficiency of the nonlinear process. In Fig.~\ref{fig:results} we show a comparison of the experimental results to a nonlinear FDTD simulation reproducing the experiment conditions  (antenna length L = 460 nm, period p = 800 nm). The linear properties of the ITO film are based on the experimental measurements shown in Fig.~\ref{fig:strongcoupling}(a), while the nonlinear response is described via the third order nonlinear susceptibility $\chi^{(3)}=9\cdot10^{-18}$ m$^2$/V$^2$ \cite{ito-chi3}. In these simulations, we blue-shift the central wavelength of the probe by 100 nm from the pump in order to discriminate the output NR and PC fields. Figures~\ref{fig:results}(a) and (b)  show the simulated near field distribution for the incident probe and the generated NR and PC fields.  The simulated NR efficiency (i.e. the NR efficiency normalised to the input probe energy, expressed in $\%$) is shown for a pump intensity of 1 GW/cm$^2$ in Fig.~\ref{fig:results}(c) (red dashed curve) and matches well to the raw experimental data  (black curve).  Figure~\ref{fig:results}(c) also shows the same FDTD simulations with detuned plasmonic antennas (i.e. antenna lengths of 250 and 650 nm) so as to be out of the strong coupling regime. The FWM efficiency drops in both cases by an order of magnitude, providing strong evidence that strong coupling is enhancing the nonlinear process.\\
 In Fig.~\ref{fig:results}(d) we show the measured NR signal efficiency for various pump powers indicated in the graph in GW/cm$^2$. The dotted black curve shows for comparison the NR efficiency at 2 GW/cm$^2$ pump power for the bare ITO film, multiplied by 1000. We see that the measured FWM efficiency of the metasurface is enhanced by more than four orders of magnitude when compared to the bare ITO. The absolute efficiency of both the NR and PC (one example shown, red-dotted line) nonlinear processes is of order $\sim1$\% over a very large bandwidth of $\sim 300$ nm. Both the NR and PC  exhibit the same trend, as expected for a deeply sub-wavelength film that is uniformly modulated at twice the probe beam frequency \cite{pendry2008time}. To verify the FWM process is primarily due to the strong coupling between the gold antennas and ITO film, and is not due solely to the gold nonlinearity, we repeated all experiments for the gold antennas deposited on a glass substrate. We find the gold antennas alone do not produce a detectable signal at the same pump powers. \\
\begin{figure}[t!]
\centering
\includegraphics[width=8cm]{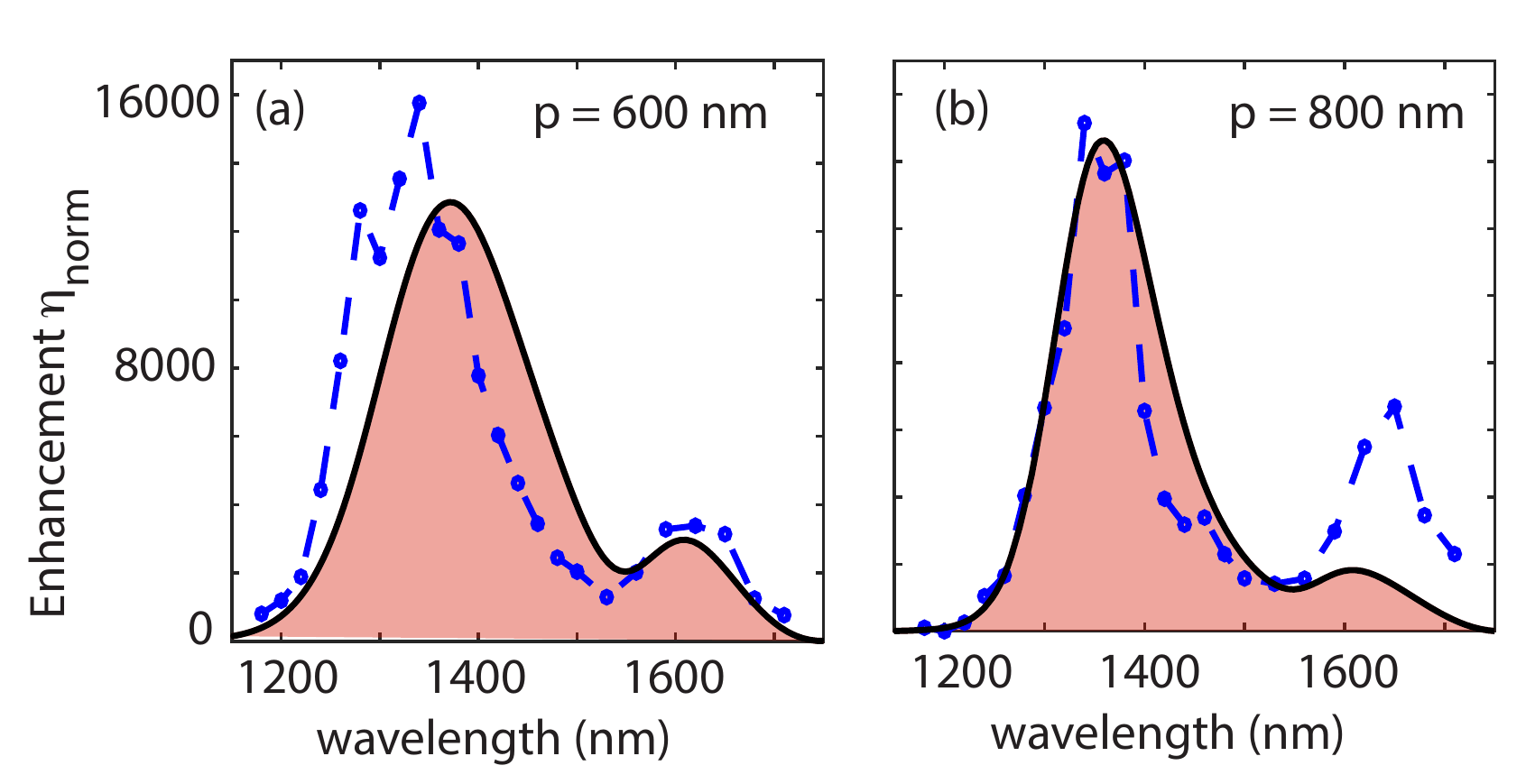}
\caption{\textbf{Theoretical model and measurements.} (a) NR efficiency $\eta_{\textrm{norm}}$ as measured for the antenna-ITO coupled system (red) for a pump intensity 0.5 GW/cm$^2$ and periodicity 600 nm  and, (b) for periodicity 800 nm (antenna length is 460 nm in both cases).}  
\label{fig:model}
\end{figure}
%
{\bf{Model and data analysis.}} In order to further understand the nonlinear enhancement, we model the pump-probe interaction in the metasurface as a  FWM process in which the four-orders of magnitude enhancement of the negative refraction and phase conjugation processes emerge as a result of the increased optical energy density in the ENZ layer. Indeed, in the strongly coupled system the E-field of both pump and probe is strongly enhanced inside the ITO layer as the plasmonic antennas convert the incident propagating waves into waves localised in their near-field. We model the nonlinear generation of beams in a FWM process starting from the numerically simulated distribution of linear fields, as described for instance in Ref.~\cite{OBrien2015}. 
FWM is driven by a nonlinear polarization in the ITO layer, $P \propto E_{\textrm{p}}^2 E_{\textrm{s}}^{*}$, where $E_{\textrm{p}}$ is the pump field, $E_{\textrm{s}}$ the probe field and $P$ is the microscopic source of the measured fields of the NR and PC beams. By making use of the reciprocity theorem, one can calculate the expected FWM E-field generated by $P$ as:
\begin{equation}
    E(\omega) \propto \int\varepsilon_0 {\chi}^{(3)}(\omega) \vdots \mathbf{E}^2_{\textrm{p}}(\omega)\mathbf{E}_{\textrm{s}}^*(\omega) \cdot \mathbf{E}_{det}(\omega) dV
\label{final} 
\end{equation} 
where the integral is calculated over the ITO volume and $E_{det}$ is the field inside the ENZ film generated by a point-source representing the detector in the far-field (see Supplementary information). 
The FWM efficiency, measured in the experiment as the ratio between the energy radiated into NR (or PC) and the incident probe energy and is thus proportional to $|E|^2$. This is true both for the bare ITO and the metasurface. Taking the ratio between the efficiency, $\eta_{\textrm{metasurface}}$, of NR for the metasurface and the efficiency, $\eta_{\textrm{ITO}}$, of NR for bare ITO removes of the spectral dependence of the
nonlinear permittivity $\chi^{(3)}$, of the linear permittivity, the sample thickness and all other constants:
\begin{equation}
\eta_{\textrm{norm}}(\omega) = \cfrac{|E_{\textrm{metasurface}}|^2}{|E_{\textrm{ITO}}|^2}
\label{efficiency}
\end{equation}
This relation directly estimates the trend of the normalised efficiency of the NR and PC processes from the energy density calculation shown in Fig.~\ref{fig:field}(c). Figure~\ref{fig:model} shows the result, together with the corresponding measured $\eta_{\textrm{norm}}$ based on the experimental data shown in Fig.~\ref{fig:results}(d) for p = 800 nm and an additional periodicity p = 600 nm.
As can be seen, our model based on the energy density enhancement in the metasurface explains the experimental results for the two different antenna configurations studied in this work, although the longer wavelength peak in the spectrum appears to have higher visibility in the experiments with respect to the theoretical model. Our primary conclusion is that the strong coupling between the plasmonic antennas and ENZ film enhances light-matter interaction and, therefore, increases the conversion efficiency of our time-varying metasurface by a factor greater than 15000. Such efficient time-varying surfaces can be obtained by optically pumping the metasurface with relatively low and readily accessible optical pumping powers of 0.5 mW, corresponding to peak intensities on the surface of 0.5 GW/cm$^2$.\\
{\bf{Conclusions.}}
Our experiments show that strongly-coupled plasmonic antenna-ENZ systems can be temporally modulated with an optical pump beam through a $\chi^{(3)}$-mediated process. Optical pumping at frequency $\omega$ induces a nonlinearity-mediated oscillation at frequency $2\omega$, which following the original predictions \cite{Trety,pendry2008time}, leads to the generation of a phase conjugate and a negative refracted beam. We observe a 15000-fold enhancement in the negative refraction and phase conjugation signals compared to the bare ENZ films. We have developed a nonlinear model which elucidates the relation between FWM enhancement and the increased energy density inside the ENZ film which arises from strong coupling.  Efficient time-varying surfaces with optically pumping at relatively low and readily accessible powers provide a route towards applications in which light is controlled by light in compact, subwavelength devices alongside a means to investigate fundamental physics, potentially including photon pair generation from deeply subwavelength systems.\\

{\bf{Acknowledgements.}}
 {{Data for this work is available for download \cite{data}}. DF acknowledges financial support from EPSRC (UK Grants EP/M009122/1 and EP/P006078/2). M.C. acknowledges the support from the United Kingdom Research and Innovation (UKRI, Innovation Fellowship EP/S001573/1). Purdue team acknowledges support by the U.S. Department of Energy, Office of Basic Energy Sciences, Division of Materials Sciences and Engineering under Award DE-SC0017717 (sample characterization), and Air Force Office of Scientific Research (AFOSR) award FA9550-18-1-0002 (numerical modeling). S.V., S.M., A.J., S.A.M. and R.S. acknowledge funding by EPSRC (EP/P033431 and EP/M013812). S.A.M. acknowledges the Lee-Lucas Chair in Physics and the DFG Cluster of Excellence Nanoscience Initiative Munich (NIM). T.H. acknowledges the Schrödinger Scholarship.\\


\providecommand{\noopsort}[1]{}\providecommand{\singleletter}[1]{#1}%
%


\end{document}


\title{Supplementary Information: Negative refraction in time-varying, strongly-coupled plasmonic antenna-ENZ system}
	
\author{V. Bruno$^{1,\dagger}$, C. DeVault$^{2,3,\dagger}$, S. Vezzoli$^{4,\dagger}$, Z. Kudyshev$^{2,3}$, T. Huq$^{4}$, S. Mignuzzi$^{4}$, A. Jacassi$^{4}$, S. Saha$^{2,3}$, Y.D. Shah$^{1}$, S.A. Maier$^{4,7}$, D.R.S. Cumming$^{6}$, A. Boltasseva$^{2,3}$, M. Ferrera$^{5}$, M. Clerici$^6$, D. Faccio$^{1,*}$, R. Sapienza$^{4,*}$, V.M. Shalaev$^{2,3,*}$}

\affiliation{$^1$ School of Physics and Astronomy, University of Glasgow,  G12 8QQ Glasgow, United Kingdom}
\affiliation{$^2$ Purdue Quantum Science and Engineering Institute, Purdue University 1205 West State Street, West Lafayette, Indiana 47907, USA}
\affiliation{$^3$ School of Electrical and Computer Engineering and Birck Nanotechnology Center, Purdue University, 1205 West State Street, West Lafayette, Indiana 47907, USA}
\affiliation{$^4$ The Blackett Laboratory, Department of Physics, Imperial College London, London SW7 2BW, United Kingdom}
\affiliation{$^5$ Institute of Photonics and Quantum Sciences, Heriot-Watt University,  EH14 4AS Edinburgh, United Kingdom}
\affiliation{$^6$ School of Engineering, University of Glasgow, G12 8LT Glasgow, United Kingdom}
\affiliation{$^7$ Chair in Hybrid Nanosystems, Faculty of Physics, Ludwig-Maxilimians-Universitat Munchen, 80799 Munchen, Germany}
\email{ daniele.faccio@glasgow.ac.uk, r.sapienza@imperial.ac.uk, shalaev@purdue.edu: $^{\dagger}$These authors contributed equally.}

\maketitle

\section{Metasurface fabrication}

 Our metasurface is based on a 500$\times$500 $\mu m$ 2D pattern of gold nano-antenna deposited on top of 40 nm thickness of ITO on a glass substrate. A bi-layer of poly methyl methacrylate (PMMA) was spin coated on top of the ITO film and then baked (first layer 40min, second layer overnight at 180 °C). The designs were patterned onto the resist using a Vistec VB6 electron beam lithography tool. A 5/30 nm layer of Ti/Au was deposited using a Plassys electron beam evaporator. After lift-off, the sample was then cleaned in acetone and isopropyl alcohol. The 40 nm thick ITO layer was purchased by Präzisions Glas $\&$ Optik.

\section{ENZ Modes of ITO Films}

\begin{figure}[h]
 	\centering
 	\includegraphics[width=12cm]{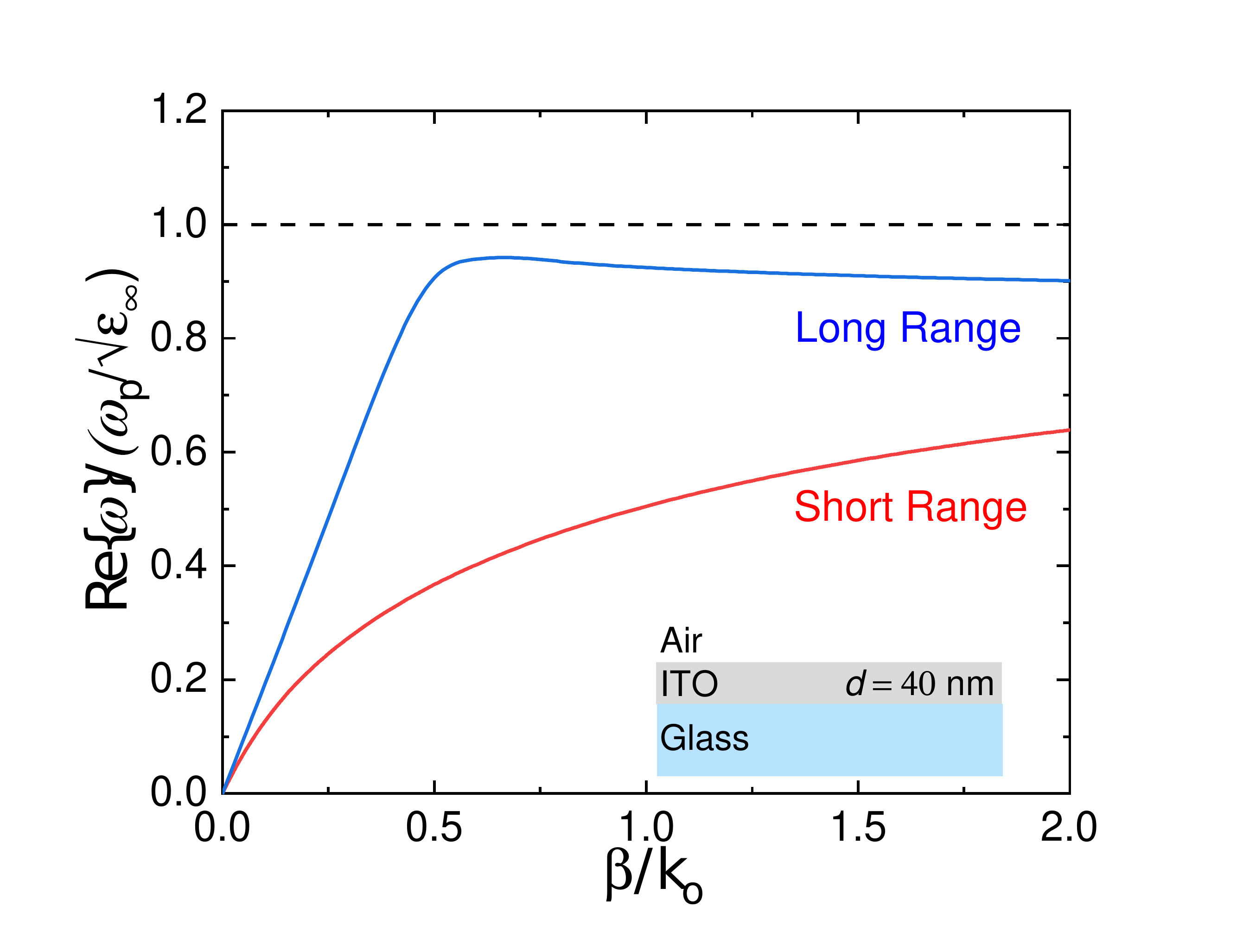}
 	\caption{Dispersion relation of the air-ENZ-glass three-layered system (see inset) for 40 nm ITO film. The long (short) range plasmon is indicated with a blue(red) line. The long range plasmon approaches the screened plasma frequency (i.e. ENZ wavelength) for large wavevectors $ \beta $, as indicated by the nearly flat dispersion.}
 	\label{fig2_modes}
\end{figure}

A thin ENZ layer supports surface modes close to the plasma frequency called an ENZ mode \cite{vassant2012ENZmodes,Vassant:12}. To find the dispersion of the ENZ mode of our bare ITO films, we solve the following dispersion equation for modes of a three layer structure
\begin{equation}
f(\beta,\omega) = 1 + \frac{\epsilon_{1}\gamma_{3}}{\epsilon_{3}\gamma_{1}} - i\tan(\kappa d)\left(\frac{\epsilon_{2}\gamma_{3}}{\epsilon_{3}\kappa}+\frac{\epsilon_{1}\kappa}{\epsilon_{2}\gamma_{1}}\right) = 0.
\label{eq:enzdisp}
\end{equation}
In Eq.~\ref{eq:enzdisp}, $\omega$ is the angular frequency, $\beta$ is the transverse wavenumber, $k_{o}^{2} = \omega^{2}/c^{2}$ is the free space wavenumber, $\gamma_{1,3} = \pm\sqrt{\epsilon_{1,3}k_{o}^{2} - \beta^{2}}$ are the longitudinal wavenumber in the superstrate ($i = 1$) and superstrate ($i = 3$), and $\kappa = \sqrt{\beta^{2} - \epsilon_{2}k_{o}^{2}}$ is the longitudinal wavenumber in the ITO layer. Here, we choose to solve Eq.~\ref{eq:enzdisp} using a real $\beta$, complex $\omega$ approach in order to capture the transient radiative decay behavior of the ENZ mode. Fig.~\ref{fig2_modes} shows solutions of Eq.~\ref{eq:enzdisp} (i.e. dispersion curves) for both long (blue lines) and short (red line) range surface plasmon  modes of the ITO thin films. It is evident that the long range mode is an ENZ mode since the dispersion is nearly flat for a large range of transverse wavevectors $ \beta $ near the screened plasma frequency (i.e. the frequency at which $\epsilon$ is close to zero). Because the long range surface modes for a 40nm ITO film occur at frequencies less than the screened plasma frequency, we expect the resonance splitting will be  red-shifted from the exact ENZ wavelength. 

\section{Coupled Harmonic Oscillator Model}

\begin{figure}[h]
	\centering
	\includegraphics[width=12cm]{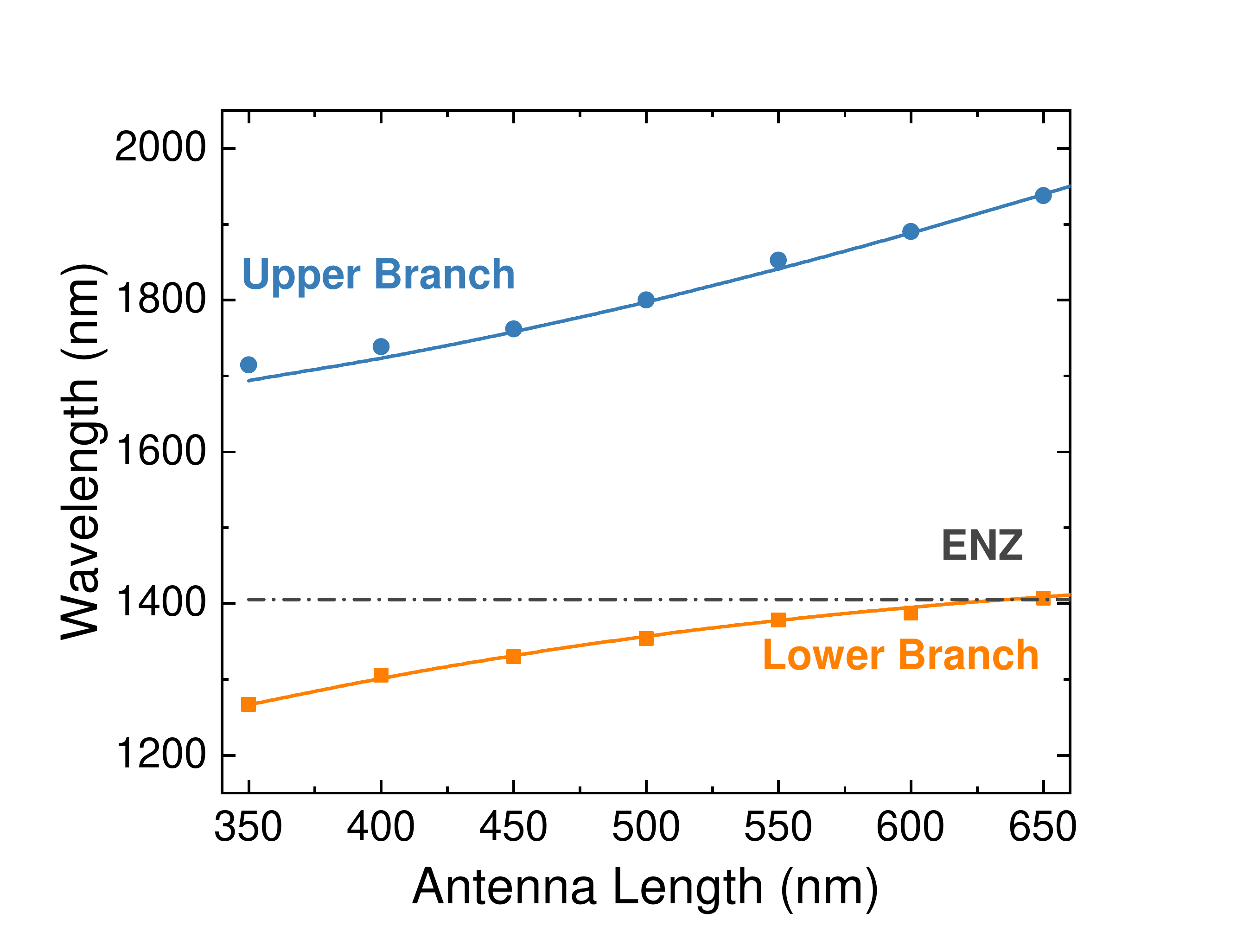}
	\caption{Wavelengths of the transmission minimum as a function of antenna length. Blue (Red) dots: minimum experimental transmission of ENZ (antenna) mode . Blue (Red) line: theoretical fit of the coupled harmonic oscillator model to the ENZ (antenna) mode. Black horizontal dashed (dashed dot) line: ENZ wavelength of the bare ITO film.}
	\label{fig4_cho}
\end{figure}

Here, we use a coupled harmonic oscillator model to retrieve the Rabi splitting of strong coupling system based on gold nanoantenna on top of deeply sub-wavelength ITO film. First,  consider two coupled harmonic oscillators and write the system of equation for the harmonic mode amplitude of oscillator one $x_{1}(\omega)$ and two $x_{2}(\omega)$ as,
\begin{equation}
\begin{pmatrix}
\omega_{1} - \omega - i\gamma_{1} & g\\ g & \omega_{2} - \omega - i\gamma_{2} 
\end{pmatrix}
\begin{pmatrix}x_{1}(\omega)\\x_{2}(\omega)\end{pmatrix} = 
i\begin{pmatrix}
f_{1}(\omega) \\ f_{2}(\omega)
\end{pmatrix}
\label{eq:CHO}
\end{equation}
where $\omega$ is the frequency, $\omega_{1,2}$ and $\gamma_{1,2}$ are the resonance frequency and damping rates of oscillator one and two, respectively, $f_{1,2}(\omega)$ are the harmonic driving forces on oscillator one and two, respectively, and \textit{g} is the coupling constant between the two oscillators. The eigenvalues of the system are solved by setting the determinant of the 2x2 matrix to zero and solving the resulting quadratic equation. The two complex eigenvalues are then given by,
\begin{equation}
\omega_{\pm} = \frac{1}{2}\left[\omega_{1}+\omega_{2} - i(\gamma_{1} +\gamma_{2})\right] \pm \frac{1}{2}\sqrt{(\omega_{1}-\omega_{2} - i(\gamma_{1} -\gamma_{2}))^{2} + 4g^{2}}.
\label{eq:eigenval}
\end{equation}
The real portion of $\omega_{\pm}$ gives the dispersion of the eigenvalues, while the imaginary portion dictates the line width \cite{torma2014strongReview}. To retrieve the coupling constant of our sample, we fit the experimental resonance minimas to the real portion of Eq.~\ref{eq:eigenval}. Figure~\ref{fig4_cho} shows the fits (solid lines) to the experimental transmission data points (circles), along with a horizontal line indicating the ENZ wavelength of the film. We extract a coupling constant of $g = 193.1$ meV, corresponding to a Rabi slitting ($\Omega_{R} = 2g$) of $\Omega_{R} = 386.2$ meV for a 40 nm ITO sample.

To verify our system is in the strong coupling regime, we compare the Rabi frequency to the average dissipation, $ \left<\gamma\right> = (\gamma_{1}+\gamma_{2})/2$, of the antenna ($\gamma_{1}$) and the ENZ films ($\gamma_{2}$). We calculate the dissipation term of the ENZ film as the loss factor extracted from the dispersion curve and the antenna from the linewidth of a Lorentzian fit of the bare antenna resonance (i.e. antenna on a simple glass substrate). We find an average dissipation of $ \left<\gamma\right> = 103.7$ meV. Indeed, in our system,  $\Omega_{R} > \left<\gamma\right> $ which quantitatively shows we are in the strong coupling regime.
%

\section{Model for the FWM interaciton}

The nonlinear generation of beams in a FWM process can be modelled from the distribution of linear fields, as described for instance by \cite{OBrien2015}. In this section we will sketch the derivation of Eq. 1 of the main text used to model the FWM process in the strongly coupled system.
The nonlinear polarization density $\mathbf{P}(\mathbf{r},\omega)$ for the FWM process leading to the emission of a negative refraction (NR) or phase conjugate (PC) beam can be written as:
\begin{equation}
	\mathbf{P}(\mathbf{r},\omega) = \epsilon_0 \bm{\chi}^{(3)}(\omega) \vdots 
	\mathbf{E}^2_{p}(\mathbf{r},\omega)\mathbf{E}_{s}^{*}(\mathbf{r},\omega)
	\label{p_nl} 
\end{equation}
The nonlinear polarization current density inside the medium
$$
\mathbf{j}(\mathbf{r},\omega)=-i\omega\mathbf{P}(\mathbf{r},\omega)
$$
is the source for the non-linear NR (or PC) field  $\mathbf{E}$ at the position of the detector, in the far-field. Such field can be estimated by means of the reciprocity theorem. The reciprocity theorem states that a source (a current density) and a detector (a field), in two distinct volumes, can be interchanged such that
$$ 
\int_{V_1} \mathbf{j}_1 \cdot \mathbf{E}_2 ~dV = \int_{V_2} \mathbf{j}_2 \cdot \mathbf{E}_1 ~dV $$
with $\mathbf{j}_1$, $\mathbf{E}_1$, $\mathbf{j}_2$, $\mathbf{E}_2$ are the current densities and the fields in the volumes $V_1$ and $V_2$ respectively. 
In our case we have:
\begin{equation}
	\int_{V_{det}} \mathbf{j}_{det} \cdot \mathbf{E} ~dV =\int_{V_{ITO}} \mathbf{j} \cdot \mathbf{E}_{det} ~dV 
	\label{rec_ito}
\end{equation}
where we introduced a dummy current density $\mathbf{j}_{det}$ in the volume $V_{det}$ at the position of the detector, which produces a (linear) field $\mathbf{E}_{det}$ in the volume of the ITO, $V_{ITO}$. A dipole current $\mathbf{j}_{det}$ in the far-field generates a plane wave travelling in the direction of NR (or PC).
$\textbf{j}_{det}$ is a point source (prop. to $\delta(\textbf{r}))$ oscillating in the direction of the generated field $E$ and we consider only the component of the electric field $E$ parallel to the direction of the antennas:

$$
{j}_{det} E =\int_{V_{ITO}} \mathbf{j} \cdot \mathbf{E}_{det} ~dV 
$$

By combining this with Eq. \ref{p_nl} we obtain the final formula of the main text:

\begin{equation}
	E(\omega) \propto \int_{V_{ITO}}  \mathbf{P}(\mathbf{r},\omega)\cdot \mathbf{E}_{det}(\mathbf{r},\omega) dV .
	\label{eff}
\end{equation}

Or by writing explicitly all the fields:

\begin{equation}
	E(\omega) \propto \int_{V_{ITO}} \epsilon_0 \bm{\chi}^{(3)}(\omega) \vdots \mathbf{E}^2_{p}(\mathbf{r},\omega)\mathbf{E}_{s}^*(\mathbf{r},\omega) \cdot \mathbf{E}_{det}(\mathbf{r},\omega) dV
	\label{final} 
\end{equation} 

The efficiency of the FWM process is proportional to the intensity of the NR field $|E|^2$. It is convenient to evaluate the ratio of the efficiency in the strongly coupled metasurface of ITO with gold antennas compared to that of bare ITO, so that the unknown spectral dependence of $\boldsymbol{\chi}^{(3)}(\omega)$ factors out:

\begin{equation}
	\eta_{norm}(\omega) = |E_{metasurface}|^2/|E_{ITO}|^2 
	\label{efficiency}
\end{equation}

Moreover by taking the ratio one can also rule out any experimental contribution to the spectral dependence, for instance due to slightly different pump intensity or collection efficiency at different wavelengths, as any variation will affect equally the NR efficiency spectra with or without antennas.

All fields in Eq. \ref{final} are calculated numerically by FDTD with plane wave illumination using Lumerical. 

\medskip
\providecommand{\noopsort}[1]{}\providecommand{\singleletter}[1]{#1}%